\documentclass{article}

\usepackage{amssymb,amsfonts,amsmath}
\usepackage{cite,enumerate,float}
\usepackage{color}
\usepackage{tikz}
\usetikzlibrary{arrows,snakes,backgrounds}
\usepackage{fancybox}
\usepackage{url}

\def\be{\begin{eqnarray}}
\def\ee{\end{eqnarray}}
\def\nn{\nonumber}

\def\MD{\hbox{Md}}
\def\qD{\hbox{qD}}

\definecolor{red}{rgb}{1,0,0}
\definecolor{orange}{rgb}{1,0.5,0}
\definecolor{violet}{rgb}{0.7,0,1}

\hoffset= - 1.0in         

\begin{document}

\title{\vspace{1.5cm}\bf\Large
Macdonald deformation of Vogel's universality\\ and link hyperpolynomials
}

\author{
Liudmila Bishler$^{b,c,d,}$\footnote{bishlerlv@lebedev.ru},
Andrei Mironov$^{b,c,d,}$\footnote{mironov@lpi.ru,mironov@itep.ru},
Alexei Morozov$^{a,c,d,}$\footnote{morozov@itep.ru}
}

\date{ }

\maketitle

\vspace{-6cm}

\begin{center}
  \hfill MIPT/TH-15/25\\
  \hfill FIAN/TD-06/25\\
  \hfill ITEP/TH-18/25\\
   \hfill IITP/TH-16/25
\end{center}

\vspace{3.5cm}

\begin{center}
 $^a$ {\small {\it MIPT, Dolgoprudny, 141701, Russia}}\\
 $^b$ {\small {\it Lebedev Physics Institute, Moscow 119991, Russia}}\\
$^c$ {\small {\it NRC ``Kurchatov Institute", 123182, Moscow, Russia}}\\
$^d$ {\small {\it Institute for Information Transmission Problems, Moscow 127994, Russia}}
\end{center}

\vspace{.1cm}

\begin{abstract}
Vogel's universality implies a unified description of the adjoint  sector of representation theory
for simple Lie algebras in terms of three parameters $\alpha,\beta,\gamma$, which are homogeneous coordinates of Vogel's plane.
Actually this is true (if at all) only for a piece of representation theory captured by knot/Chern-Simons theory,
where some irreducible representations are often undistinguishable and combined into new ``universally-irreducible" entities
(uirreps).
We consider from this point of view the recently discovered Macdonald deformation of quantum dimensions,
for which a kind of universality holds for the ADE series.
The claim is that universal are not Macdonald dimensions themselves, but their products with
Littlewood-Richardson coefficients, which themselves are functions of $q$ and $t$ in Macdonald theory.
These products are precisely what arises in knot/refined Chern-Simons theory.
Actually, we consider the simplest decomposition of adjoint square into six uirreps and obtain the universal formulas for hyperpolynomials of the Hopf link and, more generally, of the torus links $T[2,2n]$.
\end{abstract}

\bigskip

\newcommand\smallpar[1]{
  \noindent $\bullet$ \textbf{#1}
}

\section{Introduction}

Vogel's universality \cite{Vogel95,Vogel99} was a popular subject in mathematics, until the discovery
of zero divisors in Vogel's $\Lambda$-algebra (see a comprehensive review in \cite{KLS,KMS}),
which put under question a possibility of going beyond the Dynkin set of simple algebras
and their quantum deformations. This made the whole program less ambitious than it could be \cite{KMS}.
Still, for physics, this is not a restriction, but rather, in an opposite way, the Vogel universal deformation
of conformal and Chern-Simons theories, which are physical counterparts of representation theory,
is not very transparent.
Instead, for physicists the ``minor" part of Vogel's observations, which is of crucial importance, is
that the universality holds not for the ordinary irreducible representations, but for ``uirreps"
which are their peculiar combinations, indistinguishable on the entire Vogel's plane,
and decomposed differently for different simple algebras.
This is important, because this is exactly what happens in Chern-Simons, i.e. knot theory,
and irreducible representations, indistinguishable in knot theory are exactly the ones
unified into Vogel's uirreps.

Surprisingly or not, this observation gets further supported in Macdonald deformation
of the Vogel's universality, which was studied in ``refined Chern-Simons theory" \cite{AgSh1,AgSh2,AM1,R,AvMkrtString,Mane,AM2,AM3}
and is presumably relevant for description of knot hyperpolynomials \cite{AgSh1,AgSh2,DMMSS,Cher} (and, probably, super- and Khovanov-Rozansky polynomials \cite{DGR,Kh1,Kh2,Kh3,KhR,DM3}).
That is, in the recent paper \cite{BM} the problem of Vogel's universalization of Macdonald dimensions was
addressed in some detail, and we are now ready to extract interesting corollaries.
Consideration of \cite{BM} was focused on restriction of universality to the simply laced (ADE) case,
the problems with Macdonald generalization to other Dynkin diagrams are well-known \cite{KS,AM1,Mane}
and still persist.
Thus we restrict ourselves here to adjoint sector and to ADE diagrams.

In practical terms, the standard Vogel's universality says that there are typical universal quantities given on the Vogel's plane, the simple Lie algebras (or, better to say, root systems) being associated with some isolated points at just three lines in this plane. If one uses the three Vogel's parameters giving the projective plane: $\mathfrak{a}$, $\mathfrak{b}$ and $\mathfrak{c}$, the universal quantities are {\bf symmetric functions} of these parameters. One can scale all of these parameters at once with an arbitrary constant. One usually chooses one of the parameters, $\mathfrak{a}$ to be -2. Note that these parameters are usually denoted as $\alpha$, $\beta$ and $\gamma$, however, we denote them differently, since $\alpha$ is used in order to denote roots of the root system.
The Vogel's parameters for simple Lie algebras are listed in Table \ref{vogelparm}.
\begin{table}[!ht]
\centering
\begin{tabular}{|c|c|c|c|c|c|}
\hline
Root system & Lie algebra & $\mathfrak{a}$ & $\mathfrak{b}$ & $\mathfrak{c}$ & $\mathfrak{t} = \mathfrak{a}+\mathfrak{b}+\mathfrak{c}$ \\
\hline
$A_n$ & ${sl}_{n+1}$ & $-2$ & $2$ & $n+1$ & $n+1$ \\
$B_n$ & ${so}_{2n+1}$ & $-2$ & $4$ & $2n-3$ & $2n-1$ \\
$C_n$ & ${sp}_{2n}$ & $-2$ & $1$ & $n+2$ & $n+1$ \\
$D_n$ & ${so}_{2n}$ & $-2$ & $4$ & $2n-4$ & $2n-2$ \\
$G_2$ & ${g}_2$ & $-2$ & $\frac{10}{3}$ & $\frac{8}{3}$ & $4$ \\
$F_4$ & ${f}_4$ & $-2$ & $5$ & $6$ & $9$ \\
$E_6$ & ${e}_6$ & $-2$ & $6$ & $8$ & $12$ \\
$E_7$ & ${e}_7$ & $-2$ & $8$ & $12$ & $18$ \\
$E_8$ & ${e}_8$ & $-2$ & $12$ & $20$ & $30$ \\
\hline
\end{tabular}
\caption{Vogel's parameters}
\label{vogelparm}
\end{table}

The typical universal quantities are: the Chern-Simons partition function \cite{MkrtVes12,Mkrt13,KreflMkrt,M2}, the dimension \cite{Vogel99} and quantum dimension \cite{Westbury03,MkrtQDims} of the adjoint representation, eigenvalues of the second and higher Casimir operators \cite{LandMan06,MkrtSergVes,ManeIsaevKrivMkrt, IsaevProv, IsaevKriv,IsaevKrivProv} in these representations, the volume of simple Lie groups \cite{KhM},
the HOMFLY-PT knot/link polynomial colored with adjoint representation \cite{MMMuniv,MMuniv} and the Racah matrix involving the adjoint representation and its descendants \cite{MMuniv,ManeIsaevKrivMkrt, IsaevProv, IsaevKriv,IsaevKrivProv} (see also \cite{KLS}).

For study the latter quantities, one has to deal with the decomposition of irreducible representation product
\be
\mu\otimes \nu = \sum_\lambda n_{\mu\nu}^\lambda \cdot \lambda
\label{repprod}
\ee
where $\mu,\nu,\lambda$ are representations of the algebra, and $n_{\mu\nu}^\lambda$ are
the non-negative integer Littlewood-Richardson coefficients. It
has the following generalizations:
\begin{itemize}
\item{} It remains literally the same for representations of quantum group.
\item{} It survives further Macdonald deformation.
\item{} It is universal in the adjoint sector, i.e. when all the three representations
belong to powers of the adjoint representation.
However, in this case to make the statement true some irreps should be united into uirreps, and some representation for particular root systems may not enter the decomposition.
\end{itemize}

Eq.(\ref{repprod}) can be expressed in terms of characters.
For example, for the $A$ series ($sl_N$), it becomes a relation for the Schur polynomials:
\be
S_\mu\cdot S_\nu = \sum_\lambda  n_{\mu\nu}^\lambda \cdot  S_\lambda
\label{Schurprod}
\ee
The Schur polynomials are symmetric polynomials of variables $x_i$, $i=1,\ldots,N$, and $\prod_{i=1}^Nx_i=1$.
The characters are definitely can not be presented in a universal way. However, what one can do is to restrict
$x_i$ to the special locus $x_i=q^{N-2i+1}$, where $q$ is a parameter, which converts the character -- the Schur polynomial $S_\lambda$ into the quantum dimension of representation $\lambda$: $S_\lambda(x)\Big|_{x_i=q^{N-2i+1}}=\qD_\lambda$. This quantity is already universal if $\lambda$ belongs to the adjoint sector \cite{Westbury03,MkrtQDims}. A particular case of $q=1$ gives rise to ordinary dimensions.

After the Macdonald deformation, (\ref{Schurprod}) becomes
\be\label{main}
{\cal M}_\mu{\cal M}_\nu = \sum_\lambda  {\cal N}_{\mu\nu}^\lambda \cdot  {\cal M}_\lambda
\ee
where the only change is that the Macdonald Littlewood-Richardson coefficients ${\cal N}_{\mu\nu}^\lambda$ become functions of Macdonald parameters $q$ and $t$,
and they continue to vanish whenever $n_{\mu\nu}^\lambda$ vanishes, thus preserving the structure of representation products
(and description of irreps in terms of Young diagrams).
The special locus where the Macdonald polynomial factorizes is now $x_i=t^{N-2i+1}$, and one defines ``Macdonald dimension" by
\be
\MD_\mu:= {\cal M}_\mu(x)\Big|_{x_i=t^{N-2i+1}}
\ee
They play a big role in the theory of hyperpolynomials \cite{DMMSS}, and probably of superpolynomials as well \cite{DM3,AnoM}.

Similar formulas exist for other simple algebras, and
we make a conjecture that universal are the products:

\begin{center}
\fbox{\parbox{11cm}{The product
${\cal N}_{\mu\nu}^\lambda \cdot \MD_\lambda$ is universal for the triples of uirreps $\mu$, $\nu$, $\lambda$\\ from the adjoint sector, probably, in the simply laced case only.
}}
\end{center}

This conjecture is closely related to another conjecture:

\begin{center}
\fbox{\parbox{8cm}{In the simply laced case, hyperpolynomials of knots and links admit a universal representation.
}}
\end{center}

One has to point out that by universality we mean a representation in terms of the Vogel's parameters, however, the formulas are not just symmetric in $\mathfrak{a}$, $\mathfrak{b}$, $\mathfrak{c}$. Instead, {\bf they are symmetric in $u:=q^{\mathfrak{a}}$, $v:=t^{\mathfrak{b}}$, $w:=t^{\mathfrak{c}}$.
In other words, in order to make them symmetric in $\mathfrak{a}$, $\mathfrak{b}$, $\mathfrak{c}$, one has to redefine the Vogel's parameters in the table: $\mathfrak{b}\to\beta\mathfrak{b}$, $\mathfrak{c}\to\beta\mathfrak{c}$ with $t=q^\beta$.}

In this paper, we demonstrate that the first conjectures is really true in the simplest particular case of the square of adjoint representation,
i.e. when $\mu=\nu={  Adj}$, and $\lambda$ are the celebrated six Vogel's uirreps.
This is already a non-trivial calculation, especially for the $E$ series.

We also demonstrate that the second conjecture is true in the case of hyperpolynomials of the torus links $T[2,2n]$, the simplest of which is the Hopf link.

\bigskip

The paper is organized as follows. In sections 2-4, we give a detailed explanation of (\ref{main}) for ${ Adj}^{\otimes 2}$.
We begin in section \ref{ADser} with analysis of the $A$ and $D$ series.
In fact, already in this case there is subtlety,
because there are three uirreps $Y_2(\mathfrak{a})$,  $Y_2(\mathfrak{b})$, $Y_2(\mathfrak{c})$,
with the corresponding products ${\cal N}_{Adj,Adj}^{Y_2} \cdot \MD_{Y_2}$ that are {\it not} symmetric in $\mathfrak{a}$,  $\mathfrak{b}$, $\mathfrak{c}$, and the universal formula connects three different formulas by permutations.
This is true in the quantum case, but is already a non-trivial restriction on their Macdonald deformation.
{\bf Our main result of this 3 sections is (\ref{du2})} in section \ref{univformulas}, a specialization of (\ref{main}) to square of the adjoint representation. In section 4, we analyze the $E$ series in order to confirm our result. In section 5, we use the obtained results in order to obtain a universal expression for hyperpolynomials of the torus links $T[2,2n]$, which include the Hopf hyperpolynomial $T[2,2]$, and check our second conjecture.
At last, section \ref{conc} is a brief conclusion.

\paragraph{Notation and comments.}
\begin{itemize}
\item To denote the roots of the root system $R$, we use the letter $\alpha$, and we denote the set of all positive roots as $R_+$. The Macdonald polynomial associated with the root system $R$ is denoted through $P^R_\Lambda$, where $\Lambda$ is the highest weight of the representation, i.e.
\begin{equation}
    \Lambda = \sum_{i=1}^n \Lambda_i \omega_i,
\end{equation}
where $\Lambda_i \in \mathbb{Z}_{+} $, and $\omega_i$ are the fundamental weights $(\omega_i, \alpha_j^{\vee})= \delta_{ij}$, $\alpha^{\vee} = 2\alpha/(\alpha,\alpha)$. For the $A_n$ series, another notation is traditionally used: $\lambda_i=\sum_{j=i}^n\Lambda_j$, and these $\lambda_i$ give the Young diagram (partition): $\lambda_1\ge\lambda_2\ge\ldots\ge\lambda_n\ge 0$, $|\lambda|=\sum\lambda_i$.
We also sometimes use the notation ${\cal M}_\lambda$ in order to distinguish the Macdonald polynomials for the $A$ series.
\item
In variance with the original work by I. Macdonald \cite{Mac} and subsequent papers on the subject \cite{MacConj,CherednikConj,CherednikDAHA,Koorn},
we use symmetric quantum numbers, which allows us to present the results in a shorter and more elegant form: in our notation, the Macdonald polynomial depends on the squares of the original Macdonald parameters:
\begin{equation}
    q \,\rightarrow \, q^2, \,\, t \,\rightarrow \, t^2
\end{equation}
and the symmetric bracket and quantum numbers are defined to be
\begin{equation}
    \{x\} = x-x^{-1}, \quad [n]_t = \frac{t^n-t^{-n}}{t-t^{-1}}.
\end{equation}
\item When we say that some symmetric polynomial that depends on variables $x_1, \dots, x_n$ is taken at the point $q^{2\rho}$, where $\rho = (\rho_1, \dots,\rho_n)$ is the Weyl vector, we mean that one should make the substitution in this symmetric polynomial
\begin{equation}
    x_i = q^{2\rho_i}.
\end{equation}
The number of variables $x_i$ and the length of the Weyl vector coincide and are equal to the dimension of the Euclidean space where the root lattice embedded.
\end{itemize}

\section{Square of adjoint representation
\label{ADser}}

The standard decomposition of the adjoint representation square for all simple Lie groups is
\be
Adj^{\otimes 2}=X_2+Y_2(\mathfrak{a})+Y_2(\mathfrak{b})+Y_2(\mathfrak{c})+Adj+\emptyset
\label{squareadjoint}
\ee
where some of these representations can be absent or (as in the $A_n$-series case) split into two representations. The dimension and the quantum dimension of the adjoint representation admit a universal description in terms of symmetric rational functions of the Vogel's parameters \cite{Westbury03,MkrtQDims}.
Moreover, in the case of simply laced root systems, this property admits a refined ($q,t$)-generalization \cite{BM}.
A notable property of the decomposition (\ref{squareadjoint}) is that it respects the universality: the ordinary and quantum dimensions of representation $X_2$ can be realized in terms of symmetric rational functions of the Vogel's parameters, and the dimensions of $Y_2(\mathfrak{a})$, $Y_2(\mathfrak{b})$, $Y_2(\mathfrak{c})$ can be presented in a universal form, and the sum of them is also a symmetric function of $\mathfrak{a}$, $\mathfrak{b}$ and $\mathfrak{c}$.

In the next three sections, we are going to describe how these properties are extended to the refined case for the simply laced root systems, and then use this decomposition in order to construct universal knot/link hyperpolynomials.

\subsection{A series}

In the case of $A_{N-1}$ root system, and, for $N\ge 4$, the decomposition (\ref{squareadjoint}) is
\be\label{Aex2}
Adj^{\otimes 2}=[21^{N-2}]\otimes [21^{N-2}] =\underbrace{\Big([42^{N-3}1^2]=[31^{N-3}]\Big)
\oplus [332^{N-3}]}_{X_2}\oplus \underbrace{[42^{N-2}]}_{Y_2(\mathfrak{a})}\oplus \nn \\
\oplus \underbrace{\Big([332^{N-4}1^2]=[221^{N-4}]\Big)}_{Y_2(\mathfrak{b})}
\oplus 2\underbrace{\cdot\Big([32^{N-2}1]=[21^{N-2}]\Big)}_{Adj}
\oplus \Big([2^N]=[0]\Big)
\ee
i.e. there are 7 representations, and the two mutually conjugate non-adjoint representations $[42^{N-3}1^2]=[31^{N-3}]$ and $[332^{N-3}]$ form representation $X_2$. Other correspondences are
\be\label{Ac}
Adj=[21^{N-2}]\nn\\
X_2=[31^{N-3}]\oplus [332^{N-3}]\nn\\
Y_2(\mathfrak{a})=[42^{N-2}]\nn\\
Y_2(\mathfrak{b})=[332^{N-4}1^2]
\ee
Note that, for the $A$ series, the dimension of $Y_2(\mathfrak{c})$ coincides with that of the adjoint representation $[21^{N-2}]$, and one may think that the coefficient 2 in front of the adjoint representation in the decomposition (\ref{Aex2}) is because, in (\ref{squareadjoint}), there are both $Y_2(\mathfrak{c})$ and $Adj$, and they coincide. However, as we show later, universal formulas imply that, in fact, one has just to remove $Y_2(\mathfrak{c})$ from the list of representations emerging in (\ref{squareadjoint}) for the $A$ series (similarly to the exceptional root systems), and to assign the coefficient 2 with the adjoint representation in this decomposition.

The decomposition formula (\ref{squareadjoint}) can be also treated as a formula relating the corresponding characters. In particular, in the $SU(N)$ case, the characters are the Schur functions, and (\ref{squareadjoint}) is just the formula that involves the Littlewood-Richardson coefficients:
\be\label{Sd}
S_{[2,1^{N-2}]}^2=S_{[42^{N-3}1^2]}+S_{[332^{N-3}]}+S_{[42^{N-2}]}+S_{[332^{N-4}1^2]}+2S_{[32^{N-2}1]}+S_{[2^N]}\longrightarrow\nn\\
\longrightarrow S_{[31^{N-3}]}+S_{[332^{N-3}]}+S_{[42^{N-2}]}+S_{[221^{N-4}]}+2S_{[21^{N-2}]}+1
\ee
for the Schur functions $S_R(x)$ which are symmetric polynomials of variables $x_i$, $i=1,\ldots,N$, and $\prod_{i=1}^Nx_i=1$. This realization through the Schur functions is especially convenient, since, the values of these latter evaluated at the point $q^{2\rho}$, where $\rho=\frac{1}{2} \sum_{\alpha\in R_+} \alpha$ is the Weyl vector, give rise to the quantum dimensions. Thus, one immediately obtains from the decomposition (\ref{Sd}) the corresponding decomposition of quantum dimensions.

Under the $q,t$-deformation, the Schur polynomials are substituted by the Macdonald polynomials $P_\lambda^A$, counterparts of the Littlewood-Richardson coefficients $C_{\mu}^A$ become non-trivial rational functions of $q$ and $t$, and the relation takes the form
\be
\left(P^A_{[21^{N-2}]}\right)^2&=&{(t^2+1)(q^2-1)\over q^2t^2-1}P^A_{[42^{N-3}1^2]}+{(t^2+1)(q^2-1)\over q^2t^2-1}P^A_{[332^{N-3}]}+P^A_{[42^{N-2}]}+\\
&+&{(q^2-1)^2(t^2+1)^2\over (q^2t^2-1)^2}P^A_{[332^{N-4}11]}+C^A_{[32^{N-2}1]} P^A_{[32^{N-2}1]}
+{\{q\}^2\over\{t\}^2}{\{q^2A/t^2\}\{A/t\}\{A\}\{qA\}\over\{q^2A/t\}\{qA/t\}^2\{qA/t^2\}}P^A_{[2^N]}\nn
\ee
where $A:=t^N$, and, in the case of $N$ variables $x_i$, $i=1,\ldots,N$ with $\prod_{i=1}^Nx_i=1$, this formula turns into
\be\label{Aex}
\left(P^A_{[21^{N-2}]}\right)^2&=&{(t^2+1)(q^2-1)\over q^2t^2-1}P^A_{[31^{N-3}]}+{(t^2+1)(q^2-1)\over q^2t^2-1}P^A_{[332^{N-3}]}+P^A_{[42^{N-2}]}+\\
&+&{(q^2-1)^2(t^2+1)^2\over (q^2t^2-1)^2}P^A_{[221^{N-4}]}+C^A_{[21^{N-2}]} P^A_{[21^{N-2}]}
+{\{q\}^2\over\{t\}^2}{\{q^2A/t^2\}\{A/t\}\{A\}\{qA\}\over\{q^2A/t\}\{qA/t\}^2\{qA/t^2\}}\nn
\ee
Again, the two mutually conjugate non-adjoint representations $[42^{N-3}1^2]=[31^{N-3}]$ and $[332^{N-3}]$ form representation $X_2$, and the corresponding coefficients in the decomposition as well as the Macdonald dimensions coincide. The constant $C^A_{[21^{N-2}]}$ is quite involved, and we do not need its explicit form here. It
turns into 2 in (\ref{Sd}) in the non-deformed case of $t=q$.

\subsection{D series}

In the $D_n$-series case ($n\geq 5$), the list of representations is as follows:
\be
Adj=[11]\nn\\
X_2=[211]\nn\\
Y_2(\mathfrak{a})=[22]\nn\\
Y_2(\mathfrak{b})=[1111]\nn\\
Y_2(\mathfrak{c})=[2]
\ee
where we denoted representations with their highest weight in the orthogonal basis.

\noindent
The decomposition of the square of the $D$-Macdonald adjoint polynomial is:
\be\label{Dex}
\left(P^D_{[11]}\right)^2={(t^2+1)(q^2-1)\over q^2t^2-1}P^D_{[211]}+P_{[22]}^D+{\{q\}\{qt\}\{t^3\}\{t^4\}\over\{qt^2\}\{qt^3\}\{t\}\{t^2\}}P^D_{[1111]}+
{\{q\}\over\{t\}}{\{A\}\{qA/t^2\}\over\{qA/t\}\{A/t\}}{\{A^2/t^2\}\over\{qA^2/t^3\}}P^D_{[2]}+\nn\\
+C_{[11]}^DP^D_{[11]}+{\{q\}\{qt\}\over\{t\}\{t^2\}} {\{At\}\{qA/t^2\}\over\{qA\}\{A/t\}} {\{A^2\}\{A^2/t^2\}\{qA^2\}\{q^2A^2/t^2\}\over\{qA^2/t\}^2\{q^2A^2/t\}\{qA^2/t^3\}}
\ee
where $A=t^{n-1}$ for the $D_n$ case. The constant $C_{[11]}^D$ being a counterpart of $C^A_{[21^{N-2}]}$ is also involved.

\subsection{Macdonald dimensions}

Now we evaluate the decompositions (\ref{Aex}) and (\ref{Dex}) at the points $\vec x=q^{2\rho_k}$ where the refined Weyl vector
\begin{equation}
    \rho_k  = \frac{1}{2} \sum_{\alpha\in R_+} k_{\alpha} \, \alpha,
    \label{refinedWeyl}
\end{equation}
where the parameters $k_{\alpha}$ depend only on the length of the root $(\alpha,\alpha)$ (in the case of the simply laced root systems considered here, all the roots are of the same length).
The corresponding values of the Macdonald polynomials $P^R_\Lambda$ at these points are called {\bf Macdonald dimensions} $\MD^R(\Lambda)$ \cite{BM}.
The key point is that the Macdonald polynomials for any root system factorize \cite{Mac} at the point $\vec x=q^{2r_k}$, where $r_k$ is the refined version of the dual Weyl vector:
\begin{equation}\label{drW}
    r_k  = \frac{1}{2} \sum_{\alpha\in R_+} k_{\alpha} \, \alpha^{\vee},
\end{equation}
where again the parameters $k_{\alpha}$ depend only on the length of root $(\alpha,\alpha)$.
However, for the simply laced root systems, $\rho_k=r_k=k\rho$ with $t=q^k$, where $\rho$ is the ordinary Weyl vector, and, hence, the Macdonald dimensions factorize, the factorization formula being \cite{Mac} (it was later proved by I. Cherednik \cite{CherednikConj})
\begin{equation}
\MD^R(\Lambda) :=P^R_\Lambda(q^{2\rho_k})= \prod_{\alpha\in R_+} \, \prod_{j=1}^{(\alpha,\Lambda)}\,  \frac{\left\{t q^{(\rho_k,\alpha)+j-1}\right\}}{\left\{q^{(\rho_k,\alpha)+j-1}\right\}}=
\prod_{\alpha\in R_+} \, \prod_{j=1}^{(\alpha,\lambda)}\,  \frac{\left\{t^{(\rho,\alpha)+1}q^{j-1}\right\}}{\left\{t^{(\rho,\alpha)}q^{j-1}\right\}}
\end{equation}

These Macdonald dimensions in the $A$ series case are
\be
\MD^A_{Adj}={\{A/t\}\{A\}\{qA\}\over\{qA/t\}\{t\}^2}\nn\\
{1\over 2}\MD^A(X_2)={\{A/t^2\}\{A/t\}\{A\}\{qA\}\{q^2A\}\over\{q^2A/t^2\}\{qt\}\{t^2\}\{t\}^2}\nn\\
\MD^A(Y_2(\mathfrak{a}))={\{A\}\{A/t\}\{qA/t\}\{qA\}\{q^2A\}\{q^3A\}\over\{q^2A/t\}\{q^3A/t\}\{qt\}^2\{t\}^2}\nn\\
\MD^A(Y_2(\mathfrak{b}))={\{A/t^3\}\{A/t^2\}\{A/t\}\{A\}\{qA\}\{qA/t\}\over\{qA/t^3\}\{qA/t^2\}\{t^2\}^2\{t\}^2}
\ee
and, in the $D_n$-series case, are
\be
\MD^D_{Adj} = \MD^D([1,1]) = \frac{\{At\}\{A^2\}\{qA^2\}\{A^2/t^2\}}{\{t\}\{t^2\}\{A/t\}\{q A^2/t\}}\nn \\
\MD^D(X_2) = \MD^D([2,1,1]) = \frac{\{At\}\{qAt\}\{A^2\}\{A^2/t^2\}\{A^2/t^4\}\{qA^2\}\{q^2A^2\}\{qA^2/t^2\}}{\{t\}^2\{t^2\}\{qt^3\}\{A/t^2\}\{qA\}\{q^2A^2/t^2\}\{qA^2/t^3\}} \nn \\
\MD^D(Y_2(\mathfrak{a})) = \MD^D([2,2]) = \frac{\{At\}\{A^2/t^2\}\{A^2\}\{qAt\}\{qA^2/t^2\}\{qA^2\}\{q^2A^2\}\{q^3A^2\}}{\{t\}\{t^2\}\{qt\}\{qt^2\}\{A/t\}\{qA/t\}\{q^2A^2/t\}\{q^3A^2/t\}} \nn \\
\MD^D(Y_2(\mathfrak{b})) = \MD^D([1,1,1,1]) =\frac{\{At\}\{A^2\}\{A^2/t^2\}\{A^2/t^4\}\{A^2/t^6\}\{qA^2\}\{qA^2/t^2\}}{\{t\}\{t^2\}\{t^3\}\{t^4\}\{A/t^3\}\{qA^2/t^3\}\{qA^2/t^5\}} \nn \\
\MD^D(Y_2(\mathfrak{c})) = \MD^D([2]) = \frac{\{At\}\{qAt\}\{A^2\}\{qA^2\}}{\{t\}\{qt\}\{A\}\{q A\}}\nn \\
\ee

We describe the $E$ series in a separate section \ref{Eser}.

\section{The universal formulas
\label{univformulas}}

The decompositions for the Macdonald dimensions become of the form
\be\label{ex}
\left(\MD_{Adj}^R\right)^2=\sum_{\lambda}C^R_\lambda\cdot\MD^R_{\lambda}
\ee
where $\lambda=[X_2,Adj,{Y}_2(\mathfrak{a}),{Y}_2(\mathfrak{b}),{Y}_2(\mathfrak{c}),\emptyset]$, $R$ is the root system and $C_{\lambda}^R$ are $(q,t)$-deformations of Littlewood-Richardson coefficients from (\ref{Aex}) and (\ref{Dex}).

This expression can be written in a form looking universal using formulas (\ref{Aex}) and (\ref{Dex}):
\be
\left(\MD_{Adj}^R\right)^2={\xi(t^2)\over \xi(t)}\cdot\MD(X_2)+\MD(Y_2(\mathfrak{a}))+
{\xi(t^{\mathfrak{b}})\xi(t^{\mathfrak{b}/2+1})\over\xi(t)\xi(t^{\mathfrak{b}/2})}\cdot\MD(Y_2(\mathfrak{b}))+\nn\\
+{\xi(t^{\mathfrak{c}})\xi(t^{\mathfrak{c}/2+1})\over\xi(t)\xi(t^{\mathfrak{c}/2})}\cdot\MD(Y_2(\mathfrak{c}))+C_{Adj}\cdot\MD(Adj)+C_\emptyset
\ee
where
\be
\xi(x):={\{x\}\over\{qx/t\}}\stackrel{t=q}{\longrightarrow}1
\ee
However, this is not that simple to build {\bf proper}\footnote{Contributions of the representations $X_2$, $Adj$ and singlet should be
symmetric in $\mathfrak{a}$, $\mathfrak{b}$, $\mathfrak{c}$, while those of $Y_2(\mathfrak{a})$, $Y_2(\mathfrak{b})$, $Y_2(\mathfrak{c})$ should turn to each other upon permutations of $\mathfrak{a}$, $\mathfrak{b}$, $\mathfrak{c}$.} universal formulas for the Macdonald dimensions $\MD^R_{\lambda}$ entering (\ref{ex}), neither is simple to construct them for the expansion coefficients $C^R_\lambda$.
For instance, as one can see in this formula that the coefficients ${C_{{Y}_2(\mathfrak{b})}}$ and ${C_{{Y}_2(\mathfrak{c})}}$ can be presented in a proper universal form so that they turn to each other upon the permutation $\mathfrak{b}\leftrightarrow\mathfrak{c}$. At the same time, the coefficient ${C_{{Y}_2(\mathfrak{a})}}$ is just 1, and, therefore, there is no covariance in the coefficients ${C_{{Y}_2}}$, when permuting $\mathfrak{a}$ with $\mathfrak{b}$ and $\mathfrak{c}$. However, the whole combinations with the Macdonald dimensions are already covariant: $\mathfrak{Y}_2(\mathfrak{a}):=C_{Y_2(\mathfrak{a})}\cdot \MD_{Y_2(\mathfrak{a})}$ transforms to $\mathfrak{Y}_2(\mathfrak{b}):=C_{Y_2(\mathfrak{b})}\cdot \MD_{Y_2(\mathfrak{b})}$ at $\mathfrak{a}\to\mathfrak{b}$, etc.

Thus, for the combinations $C^R_\lambda\cdot\MD^R_{\lambda}$, it is possible to construct universal formulas! Since, for constructing universal knot/link invariants, one needs only universality of the whole terms, not of separate multipliers, this opens an immediate way to build these universal invariants.

Thus, the universal decomposition formula looks like
\be\label{du1}
\boxed{
\left(\MD_{Adj}\right)^2=\mathfrak{X}_2+\mathfrak{Y}_2(\mathfrak{a})+\mathfrak{Y}_2(\mathfrak{b})+\mathfrak{Y}_2(\mathfrak{c})
+\mathfrak{P}_{Adj}+\mathfrak{P}_\emptyset
}
\ee
where, in the variables
\be
u:=q^{\mathfrak{a}},\ \ \ \ \ v:=t^{\mathfrak{b}},\ \ \ \ \ w:=t^{\mathfrak{c}},\ \ \ \ \ T:={q^2\over t^2}uvw\nn
\ee
the separate terms of the decomposition are

\bigskip

\doublebox{\parbox{16cm}{
\be\label{du2}
    \MD_{Adj}&=&-{\left\{\frac{T}{\sqrt{u}}\right\}\left\{\frac{T}{\sqrt{v}}\right\}\left\{\frac{T}{\sqrt{w}}\right\}\over
    \{\sqrt{u}\}
    \{\sqrt{v}\}\{\sqrt{w}\}}\cdot{\xi(T)\over \xi(t)}
    \nn\\ \nn \\
    \mathfrak{X}_2&=&-\MD_{Adj}\times{\left\{\frac{q}{t}\sqrt{Tu}\right\}\left\{\frac{q}{t}\sqrt{Tv}\right\}
    \left\{\frac{q}{t}\sqrt{Tw}\right\}\left\{\frac{T}{u}\right\}\left\{\frac{T}{v}\right\}\left\{\frac{T}{w}\right\}\over
    \left\{\frac{qu}{t}\right\}\left\{\frac{qv}{t}\right\}\left\{\frac{qw}{t}\right\}
    \left\{\sqrt{\frac{T}{u}}\,\right\}\left\{\sqrt{\frac{T}{v}}\,\right\}\left\{\sqrt{\frac{T}{w}}\,\right\}}
   \cdot
      {\xi(qT)\xi\left(t\sqrt{T}\right)\xi\left(\frac{\sqrt{T}}{t}\right)\over
    \xi(t^2T)\xi(qtT)}\cdot{\xi(t^2)\over\xi(t)\xi\left(\frac{1}{t^2}\right)}
    \nn\\ \nn \\
\mathfrak{Y}_2(\mathfrak{a})&=&\MD(Y_2(\mathfrak{a}))=
{\left\{\frac{T}{u\sqrt{u}}\right\}\left\{\frac{T}{\sqrt{uv}}\right\}\left\{\frac{T}{\sqrt{uw}}\right\}
\left\{\frac{T}{\sqrt{v}}\right\}\left\{\frac{T}{\sqrt{w}}\right\}\{T\}
\over
\left\{ u \right\} \left\{\sqrt{ u} \right\}\left\{\sqrt{ v} \right\}\left\{\sqrt{ w} \right\}\left\{\sqrt{\frac{v}{ u}} \right\}\left\{\sqrt{ \frac{w}{u}} \right\}}
\cdot \frac{\xi\left(\frac{T}{{u}}\right) \xi\left(\frac{T}{\sqrt{u}}\right)\xi\left(u\right)  }{\xi\left(t\right)^2\xi\left(\sqrt{u}\right)}
\nn\\
\mathfrak{Y}_2(\mathfrak{b})&=&
C_{Y_2(\mathfrak{b})}\cdot \MD(Y_2(\mathfrak{b}))= \mathfrak{Y}_2(\mathfrak{a})\Big|_{u\leftrightarrow v}
\nn\\
\mathfrak{Y}_2(\mathfrak{c})&=&
C_{Y_2(\mathfrak{c})}\cdot \MD(Y_2(\mathfrak{c}))= \mathfrak{Y}_2(\mathfrak{a})\Big|_{u\leftrightarrow w}
\nn\\
\mathfrak{P}_\emptyset&=&{\xi\left(\frac{1}{q}\right)\over\xi(t)}\cdot {\xi(T)
\xi\left(\frac{T}{\sqrt{u}}\right) \xi\left(\frac{T}{\sqrt{v}}\right)\xi\left(\frac{T}{\sqrt{w}}\right)\over
\xi(qT/t)
\xi(\sqrt{u})\xi(\sqrt{v})\xi(\sqrt{w})}
\ee}}

\bigskip

All these quantities but $\mathfrak{X}_2$ are written in the form of products of the standard universal quantum dimensions (upon parametrization $u=q^{\mathfrak{a}}$, $v=q^{\mathfrak{b}}$, $w=q^{\mathfrak{c}}$, $T=q^{\mathfrak{a}+\mathfrak{b}+\mathfrak{c}}$ \cite{MMMuniv}) and of $\xi$-factors, which drops out at $t=q$. The expansion (\ref{du1}) is
\fbox{symmetric w.r.t. permutations of $u$, $v$, $w$.}

\bigskip

The remaining term $\mathfrak{P}_{Adj}=C_{Adj}\cdot \MD_{Adj}$, $\MD_{Adj}$ being ($\mathfrak{a}$,$\mathfrak{b}$,$\mathfrak{c}$)-symmetric, and such being also the coefficient $C_{Adj}$. However, $C_{Adj}$ is quite involved, and one
can restore $\mathfrak{P}_{Adj}$ in the universal form in the simplest way from formulas (\ref{du1})-(\ref{du2}).

Note that, for the $A_n$-series, $\mathfrak{Y}_2(\mathfrak{c})=0$, in accordance with what we explained after formula (\ref{Ac}). Moreover, this representation ${Y}_2(\mathfrak{c})$ also does not emerge in the decomposition of the adjoint representation square for the exceptional algebras as we explain in the next section.

\section{E series
\label{Eser}}

Remarkably, these formulas are also consistent with the Macdonald dimensions of the $E$-series: expansions of the adjoint representation square along with the Macdonald dimensions are, in this case, as follows.

\subsection{Adjoint square decomposition for $E_6$}
In the basis of fundamental weights $\omega_i$,
\begin{equation}
 \omega_{Adj}^{E_6} = \omega_6 
\end{equation}
and
\begin{equation}
    \textbf{78} \otimes \textbf{78} = \textbf{1} \oplus \textbf{78} \oplus \textbf{650}\oplus \textbf{2430}\oplus\textbf{2925}
\end{equation}
where
\begin{equation}
    \begin{aligned}
   Adj = \textbf{78} & \quad\quad    \omega_{Adj} = \omega_6 \\
    X_2 = \textbf{2925} & \quad\quad   \omega_{X_2} = \omega_3   \\
    Y_2(\mathfrak{a}) = \textbf{2430} & \quad\quad     \omega_{Y_2(\mathfrak{a})} = 2 \omega_6\\
     Y_2(\mathfrak{b}) = \textbf{650} & \quad\quad     \omega_{Y_2(\mathfrak{b})} = \omega_1+\omega_5 \\
     Y_2(\mathfrak{c}) = \textbf{0} & \quad\quad    
    \end{aligned}
\end{equation}

The corresponding Macdonald dimensions are
\begin{align}
    \MD^{E_6}_{Adj} = \MD^{E_6}(\omega_6) = \frac{[8]_t[9]_t[12]_t}{[3]_t[4]_t}\frac{\{q t^{12}\}}{\{q t^{11}\}} \nn\\
     \MD^{E_6}(X_2) = \MD^{E_6}(\omega_3) = \frac{[5]_t[6]_t[8]_t[9]_t[12]_t}{[2]_t^2 [3]_t^2} \frac{\{q t^8\}\{q t^9\}\{q t^{12}\}}{\{q t^5\}\{q t^6\}\{q t^7\}} \frac{\{q^2 t^{12}\}}{\{q^2 t^{10}\}} \nn\\
     \MD^{E_6}(Y_2(\mathfrak{a})) = \MD^{E_6}(\omega_6) = \frac{[8]_t[9]_t[12]_t}{[3]_t[4]_t}\frac{\{q t^{8}\}\{q t^{9}\}\{q t^{12}\}}{\{q t\}\{q t^{3}\}\{q t^{4}\}} \frac{\{q^2 t^{12}\}\{q^3 t^{12}\}}{\{q^2 t^{11}\}\{q^3 t^{11}\}}  \nn \\
     \MD^{E_6}(Y_2(\mathfrak{b})) = \MD^{E_6}(\omega_1+\omega_5)  = \frac{[5]_t[8]_t[9]_t[12]_t}{[4]_t^2} \frac{\{q t^9\}\{q t^{12}\}}{\{q t^5\}\{q t^8\}}
\end{align}

\subsection{Adjoint square decomposition for $E_7$}
In the basis of fundamental weights,
\begin{equation}
 \omega_{Adj}^{E_7} = \omega_1
\end{equation}
and
\begin{equation}
    \textbf{133} \otimes \textbf{133} = \textbf{1} \oplus \textbf{133} \oplus \textbf{1539}\oplus \textbf{7371}\oplus\textbf{8645}
\end{equation}
where
\begin{equation}
    \begin{aligned}
   Adj = \textbf{133} & \quad\quad    \omega_{Adj} = \omega_1 \\
    X_2 = \textbf{8645} & \quad\quad   \omega_{X_2} = \omega_2   \\
    Y_2(\mathfrak{a}) = \textbf{7371} & \quad\quad     \omega_{Y_2(\mathfrak{a})} = 2 \omega_1\\
     Y_2(\mathfrak{b}) = \textbf{1539} & \quad\quad     \omega_{Y_2(\mathfrak{b})} = \omega_5 \\
     Y_2(\mathfrak{c}) = \textbf{0} & \quad\quad     
    \end{aligned}
\end{equation}

The corresponding Macdonald dimensions are
\begin{align}
     \MD^{E_7}_{Adj} = \MD^{E_7}(\omega_1) = \frac{[12]_t[14]_t[18]_t}{[4]_t[6]_t} \frac{\{q t^{18}\}}{\{q t^{17}\}}  \nn\\
    \MD^{E_7}(X_2) = \MD^{E_7}(\omega_2) = \frac{[8]_t[10]_t[12]_t[14]_t[18]_t}{[2]_t[3]_t[4]_t[5]_t} \frac{\{qt^{12}\}\{qt^{14}\}\{qt^{18}\}}{\{qt^{7}\}\{qt^{9}\}\{qt^{11}\}} \frac{\{q^2 t^{18}\}}{\{q^2 t^{16}\}} \nn\\
    \MD^{E_7}(Y_2(\mathfrak{a})) = \MD^{E_7}(2\omega_1) = \frac{[12]_t[14]_t[18]_t}{[4]_t[6]_t} \frac{\{qt^{12}\}\{qt^{14}\}\{qt^{18}\}}{\{qt\}\{qt^{4}\}\{qt^{6}\}} \frac{\{q^2t^{18}\}\{q^3t^{18}\}}{\{q^2t^{17}\}\{q^3t^{17}\}}\nn\\
    \MD^{E_7}(Y_2(\mathfrak{b})) = \MD^{E_7}(\omega_5) = \frac{[10]_t[12]_t[14]_t[18]_t}{[2]_t[4]_t[5]_t} \frac{\{qt^{14}\}\{qt^{18}\}}{\{qt^{9}\}\{qt^{13}\}}
\end{align}

\subsection{Adjoint square decomposition for $E_8$}
In the basis of fundamental weights,
\begin{equation}
 \omega_{Adj}^{E_8} = \omega_7
\end{equation}
and
\begin{equation}
    \textbf{248} \otimes \textbf{248} = \textbf{1} \oplus \textbf{248} \oplus \textbf{3875}\oplus \textbf{27000}\oplus\textbf{30380}
\end{equation}
where
\begin{equation}
    \begin{aligned}
   Adj = \textbf{248} & \quad\quad    \omega_{Adj} = \omega_7 \\
    X_2 = \textbf{30380} & \quad\quad   \omega_{X_2} = \omega_6   \\
    Y_2(\mathfrak{a}) = \textbf{27000} & \quad\quad     \omega_{Y_2(\mathfrak{a})} = 2 \omega_7\\
     Y_2(\mathfrak{b}) = \textbf{3875} & \quad\quad     \omega_{Y_2(\mathfrak{b})} = \omega_1 \\
     Y_2(\mathfrak{c}) = \textbf{0} & \quad\quad     
    \end{aligned}
\end{equation}

The corresponding Macdonald dimensions are
\begin{align}
    \MD^{E_8}_{Adj} = \MD^{E_8}(\omega_7) = \frac{[20]_t[24]_t[30]_t}{[6]_t[10]_t} \frac{\{q t^{30}\}}{\{q t^{29}\}}\nn\\
\MD^{E_8}(X_2) = \MD^{E_8}(\omega_6) = \frac{[14]_t[18]_t[20]_t[24]_t[30]_t}{[2]_t[5]_t[6]_t[9]_t} \frac{\{q t^{20}\}\{q t^{24}\}\{q t^{30}\}}{\{q t^{11}\}\{q t^{15}\}\{q t^{19}\}}\frac{\{q^2 t^{30}\}}{\{q^2 t^{28}\}}\nn\\
    \MD^{E_8}(Y_2(\mathfrak{a})) = \MD^{E_8}(2\omega_7) = \frac{[20]_t[24]_t[30]_t}{[6]_t[10]_t} \frac{\{q t^{20}\}\{q t^{24}\}\{q t^{30}\}}{\{q t\}\{q t^{6}\}\{q t^{10}\}} \frac{\{q^2 t^{30}\}\{q^3 t^{30}\}}{\{q^2 t^{29}\}\{q^3 t^{29}\}}\nn\\
    \MD^{E_8}(Y_2(\mathfrak{b})) = \MD^{E_8}(\omega_1) = \frac{[14]_t[18]_t[20]_t[24]_t[30]_t}{[4]_t[6]_t[7]_t[10]_t} \frac{\{q t^{24}\}\{q t^{30}\}}{\{q t^{17}\}\{q t^{23}\}}
\end{align}

\subsection{Unified formulas for $E$ series}

In fact, after introducing the variables
\be
u = \frac{1}{q^2},\  v = t^{2s+4}, \ w = t^{4s+4}, \  T = t^{6(s+1)}
\ee
where $E_6, E_7$ and $E_8$ correspond to $s=1,2,4$, these formulas for particular exceptional groups can be unified into
\be
    \MD^{E}_{Adj} = {\{t^{4s+4}\}\{t^{5s+4}\}\{t^{6s+6}\}\over\{t\}\{t^{s+2}\}\{t^{2s+2}\}}\cdot{\{qt^{6s+6}\}\over\{qt^{6s+5}\}}  \nn\\
\MD^{E}(X_2) =  {\{t^{3s+2}\}\{t^{4s+2}\}\{t^{4s+4}\}\{t^{5s+4}\}\{t^{6s+5}\}\{t^{6s+6}\}\over\{t\}\{t^2\}\{t^{s+1}\}\{t^{s+2}\}\{t^{2s+1}\}}
\cdot{\{qt^{4s+4}\}\{qt^{5s+4}\}\over\{qt^{2s+3}\}\{qt^{3s+3}\}\{qt^{4s+3}\}}\cdot{\{q^2t^{6s+6}\}\over\{q^2t^{6s+4}\}}\nn\\
    \MD^{E }(Y_2(\mathfrak{a})) ={\{t^{4s+4}\}\{t^{5s+4}\}\{t^{6s+6}\}\over\{t\}\{t^{s+2}\}\{t^{2s+2}\}}\cdot{\{qt^{4s+4}\}\{qt^{5s+4}\}\{qt^{6s+6}\}
    \over\{qt\}\{qt^{s+2}\}\{qt^{2s+2}\}}\cdot{\{q^2t^{6s+6}\}\over\{q^2t^{6s+5}\}}\cdot{\{q^3t^{6s+6}\}\over\{q^3t^{6s+5}\}}\nn\\
    \MD^{E }(Y_2(\mathfrak{b})) ={\{t^{3s}\}\{t^{3s+2}\}\{t^{4s+2}\}\{t^{4s+4}\}\{t^{5s+4}\}\{t^{6s+6}\}\over\{t\}\{t^s\}\{t^{s+2}\}\{t^{s+3}\}\{t^{2s+2}\}
    \{t^{2s+4}\}}\cdot{\{qt^{5s+4}\}\{qt^{6s+6}\}\over\{qt^{4s+1}\}\{qt^{5s+3}\}}
\label{Eflas}
\ee
This is what we can now use in comparison with (\ref{du2}),
and there is an exact match.
Thus (\ref{du2}), i.e. i.e. our main claim from the Introduction, for the particular $\mu=\nu=Adj$ {\bf is fully proved} for all ADE algebras.

\section{Universal hyperpolynomial for torus links $T[2,2n]$}

In this section, we are going to use formula (\ref{du1}) in order to construct hyperpolynomials of the torus links $T[2,2n]$ with both link components colored with the adjoint representation. We start with the simplest example of the $T[2,2]$ Hopf link.

\subsection{Hopf hyperpolynomial for $A$ series}

In the case of $A$ series, the refined Hopf link invariant, i.e. Hopf hyperpolynomial can be defined both by the formula \cite{IK,AKMM2}
\be\label{Hopf}
{\cal P}_{\lambda,\mu}^{\rm Hopf} =\MD^A_\lambda\cdot {\cal M}_\mu(q^{2\mu} t^{2\rho})
\ee
and by the refined version \cite{AgSh1,AgSh2,DMMSS,Cher} of the Rosso-Jones formula \cite{RJ,China}
\be\label{RJ}
{\cal P}_{\lambda,\mu}^{\rm Hopf} =f_\lambda^2 f^2_\mu
\sum_{\eta\in \lambda\otimes \mu}
{\cal N}_{\lambda\mu}^\eta f_\eta^{-2}
\cdot {\MD}^A_\eta
\ee
Here
\be\label{fr}
f_\mu:=\left(-{q\over t}\right)^{|\mu|/2}q^{\nu'(\mu)/2}t^{-\nu(\mu)/2}
\ee
is the Taki framing factor \cite{Taki} and
\be\label{nu}
\nu(\lambda):=2\sum_i (i-1)\lambda_i,\ \ \ \ \ \nu'(\lambda):=\nu(\lambda^\vee)
\ee
The definition (\ref{Hopf}) coincides with (\ref{RJ}) \cite{AKMM2}. One can check that ${\cal P}_{\lambda,\mu}^{\rm Hopf}={\cal P}_{\mu,\lambda}^{\rm Hopf}$ as it should be. A discussion of the framing factor in these formulas can be found in \cite{AKMM2}.

Now, we are interested in the hyperpolynomial ${\cal P}_{Adj,Adj}^{\rm Hopf}$ and note that one can choose $f_\mu$ just in the form
\be
f_\mu:=q^{\nu'(\mu)/2}t^{-\nu(\mu)/2}
\ee
since the $q/t$-factor drops out the hyperpolynomial at $\lambda=\nu$. This factor can be written for any simply laced root system in the form (which is a refined version of exponential of the eigenvalue of the second Casimir operator)
\be
f_{\Lambda}^R=q^{(\Lambda,\Lambda)/2+(\Lambda,\rho_k)}
\ee
where $\Lambda$ is the highest weight vector giving the representation.

\subsection{Universal link hyperpolynomials}

\paragraph{Taki factor for other root systems.}
For representations emerging in (\ref{du1}), one gets the following expressions:

\paragraph{$A$ series.}
\begin{align}
   & f_{Adj}^{A_n} =  f_{[2,1^{n-1}]}^{A_n} = q t^n \nn\\
   & f_{X_2}^{A_n} =  f_{[31^{n-2}]}^{A_n}=f_{[322^{n-2}]}^{A_n}  = q^3 t^{2n-1}\nn\\
   & f_{Y_2(\mathfrak{a})}^{A_n} =  f_{[4,2^{n-1}]}^{A_n} = q^4 t^{2n} \nn\\
   & f_{Y_2(\mathfrak{b})}^{A_n} =  f_{[2,2,1^{n-3}]}^{A_n} = q^2 t^{2 n - 2}
\end{align}
\paragraph{$D$-series.}
\begin{align}
   & f_{Adj}^{D_n} =  f_{[1,1]}^{D_n} = q t^{2 n - 3}\nn\\
   & f_{X_2}^{D_n} =  f_{[2,1,1]}^{D_n}  = q^3 t^{4 n - 7}\nn\\
   & f_{Y_2(\mathfrak{a})}^{D_n} =  f_{[2,2]}^{D_n} = q^4 t^{4 n - 6}  \nn\\
   & f_{Y_2(\mathfrak{b})}^{D_n} =  f_{[1,1,1,1]}^{D_n} = q^2 t^{4 n - 10}   \nn\\
   & f_{Y_2(\mathfrak{c})}^{D_n} =  f_{[2]}^{D_n} = q^2 t^{2 n - 2}
\end{align}
\paragraph{$E$-series.}
\begin{align}
     & f_{Adj}^{E_6} = f_{\omega_6}^{E_6}= qt^{11} &&  f_{Adj}^{E_7} = f_{\omega_1}^{E_7}= q t^{17} && f_{Adj}^{E_8} = f_{\omega_7}^{E_8}= q t^{29} \nn\\
   & f_{X_2}^{E_6} = f_{\omega_3}^{E_6} = q^3 t^{21}&& f_{X_2}^{E_7} = f_{\omega_2}^{E_7} = q^3 t^{33} && f_{X_2}^{E_8} = f_{\omega_6}^{E_8} = q^3 t^{57}\nn\\
   & f_{Y_2(\mathfrak{a})}^{E_6} =  f_{2\omega_6}^{E_6} = q^4 t^{22}  && f_{Y_2(\mathfrak{a})}^{E_7} =  f_{2\omega_1}^{E_7} = q^4 t^{34} && f_{Y_2(\mathfrak{a})}^{E_8} =  f_{2\omega_7}^{E_8} = q^4 t^{58} \nn\\
   & f_{Y_2(\mathfrak{b})}^{E_6} =  f_{\omega_1+\omega_5}^{E_6} = q^2 t^{16} \quad \quad && f_{Y_2(\mathfrak{b})}^{E_7} =  f_{\omega_5}^{E_7} = q^2 t^{26} \quad \quad && f_{Y_2(\mathfrak{b})}^{E_8} =  f_{\omega_1}^{E_8} =  q^2 t^{46}
\end{align}


\bigskip
\paragraph{Universal Taki factor.}
This implies the universal expressions:

\bigskip

\doublebox{\parbox{15cm}{
\begin{align}
   & \mathfrak{f}_{Adj} = \frac{q}{t}T\nn\\
   & \mathfrak{f}_{X_2}  = \frac{q^3}{t^3} T^2\nn  \\
   & \mathfrak{f}_{Y_2(\mathfrak{a})} = {q^2\over t^2}{T^2\over u}\nn \\
   & \mathfrak{f}_{Y_2(\mathfrak{b})} = {q^2\over t^2}{T^2\over v}\nn \\
   & \mathfrak{f}_{Y_2(\mathfrak{c})} =  {q^2\over t^2}{T^2\over w} \nn\\
   & \mathfrak{f}_{\varnothing} = 1\nn
\end{align}}}

\bigskip

One can see that these expressions have proper forms, in particular, $\mathfrak{f}_{Adj}$ and $\mathfrak{f}_{X_2}$ are symmetric in $u$, $v$, $w$, and $\mathfrak{f}_{Y_2(\mathfrak{a})}$, $\mathfrak{f}_{Y_2(\mathfrak{a})}$, $\mathfrak{f}_{Y_2(\mathfrak{a})}$ turn into each other under permutations of $u$, $v$, $w$. One can also see again that the universality is preserved provided the representation $Y_2(\mathfrak{c})$ does not emerge in the decomposition (\ref{squareadjoint}) and $\mathfrak{Y}_2(\mathfrak{c})=0$ for the $A$ series: the value of $\mathfrak{f}_{Y_2(\mathfrak{c})}$ is ${q\over t}f_{Adj}$ in this case, i.e. do not coincide with that of the adjoint representation, which would breaks the universality down.

\paragraph{Universal Hopf hyperpolynomial.}
Thus, we are ready to immediately construct the universal Hopf hyperpolynomial:
\be\label{Hopfu}
\boxed{
\mathfrak{H}_{Adj,Adj}^{\rm Hopf} =\mathfrak{f}_{Adj}^4\left[\mathfrak{f}_{X_2}^{-2}\mathfrak{X}_2+\mathfrak{f}_{Y_2(\mathfrak{a})}^{-2}\mathfrak{Y}_2(\mathfrak{a})
+\mathfrak{f}_{Y_2(\mathfrak{b})}^{-2}\mathfrak{Y}_2(\mathfrak{b})+\mathfrak{f}_{Y_2(\mathfrak{c})}^{-2}\mathfrak{Y}_2(\mathfrak{c})
+\mathfrak{f}_{Adj}^{-2}\mathfrak{P}_{Adj}+\mathfrak{P}_\emptyset\right]}
\ee

Note that this quantity is divisible by $\MD_{Adj}$ and, hence, due to (\ref{Hopf}), provides {\bf a universal form for the quantity $P^R_{Adj}(q^{2\lambda_{Adj}} t^{2\rho})$}, which generalizes the adjoint Macdonald dimension
$\MD^R_{Adj}=P^R_{Adj}(t^{2\rho})$, and is no longer factorized:
\be
\boxed{
P^R_{Adj}(q^{2\lambda_{Adj}} t^{2\rho})={\mathfrak{H}_{Adj,Adj}^{\rm Hopf}\over\MD_{Adj}}
}
\ee

\paragraph{2-strand torus link hyperpolynomials.}
In order to construct hyperpolynomials for the whole series of the torus 2-strand links $T[2,2n]$, one can use the evolution method \cite{DMMSS,MMMevo}. Then, it is sufficient just to raise powers of $f_{\Lambda}$ accordingly in order to get the universal formula for the torus $T[2,2n]$ link hyperpolynomials:

\bigskip

\doublebox{\parbox{16.2cm}{
\be\label{linku}
\mathfrak{H}_{Adj,Adj}^{T[2,2n]} =\mathfrak{f}_{Adj}^{4n}\left[\mathfrak{f}_{X_2}^{-2n}\mathfrak{X}_2
+\mathfrak{f}_{Y_2(\mathfrak{a})}^{-2n}\mathfrak{Y}_2(\mathfrak{a})+
\mathfrak{f}_{Y_2(\mathfrak{b})}^{-2n}\mathfrak{Y}_2(\mathfrak{b})+\mathfrak{f}_{Y_2(\mathfrak{c})}^{-2n}\mathfrak{Y}_2(\mathfrak{c})
+\mathfrak{f}_{Adj}^{-2n}\mathfrak{P}_{Adj}+\mathfrak{P}_\emptyset\right]
\ee
}}

\bigskip

This formula provides an immediate $(q,t)$-deformation of the Rosso-Jones formula \cite{RJ,China} in the spirit of \cite{AgSh1,AgSh2,DMMSS,Cher} and of formula (\ref{RJ}). It can be also written in the form which does not involve the complicated item $\mathfrak{P}_{Adj}$:
\be
\mathfrak{H}_{Adj,Adj}^{T[2,2n]} =\mathfrak{f}_{Adj}^{4n}\left[\Big(\mathfrak{f}_{X_2}^{-2n}-\mathfrak{f}_{Adj}^{-2n}\Big)\mathfrak{X}_2
+\Big(\mathfrak{f}_{Y_2(\mathfrak{a})}^{-2n}-\mathfrak{f}_{Adj}^{-2n}\Big)\mathfrak{Y}_2(\mathfrak{a})+
\Big(\mathfrak{f}_{Y_2(\mathfrak{b})}^{-2n}-\mathfrak{f}_{Adj}^{-2n}\Big)\mathfrak{Y}_2(\mathfrak{b})
+\Big(\mathfrak{f}_{Y_2(\mathfrak{c})}^{-2n}-\mathfrak{f}_{Adj}^{-2n}\Big)\mathfrak{Y}_2(\mathfrak{c})+\right.\nn\\
+\left.\mathfrak{f}_{Adj}^{-2n}\MD_{Adj}^2+\Big(1-\mathfrak{f}_{Adj}^{-2n}\Big)\mathfrak{P}_\emptyset\right]
\ee
The crucial point is that at $n=0$, when the link becomes trivial, this formula turns back into (\ref{du1}). Note that, in the case of links, in variance with the knots there are no correcting $\gamma$-factors \cite{DMMSS}.

\subsection{Knot hyperpolynomials}

However, for knots the situation is more interesting.
Now the evolution series for $[2,2n+1]$ terminates at the trivial point $T[2,1]=T[1,2]$,
and a counterpart of (\ref{du1}) to coincide with is the adjoint Macdonald dimension $\MD_{Adj}$ instead of its square.
However, in order to get this matching, one now requires to introduce additional $\gamma$-factors \cite{DMMSS}.
For ordinary representations they can be made independent of $A$, and can be sometime restored just from
the boundary conditions for the evolution in $n$.
For the adjoint sector, however, this is not obligatory true: an $A$-dependence can arise, because Young diagrams
can explicitly depend on $N$.
At the same time, they do depend for the $A$ series, but do not for the $D$ series, which would put the universality of
$\gamma$-factors under question.
In this subsection, we provide some technical details, which can be useful for the study of these problems,
but the final answer remains obscure.

For the $A$ series, in accordance with the general rule \cite{DMMSS}, one has to consider 2-plethysm and evaluate the corresponding Adams coefficients, i.e. make a substitution $x_i\to x_i^2$ in the adjoint Macdonald polynomial $P^A_{Adj}$ and then expand to the same five Macdonald polynomials (of variables $x_i$) $P^A_X$, $P^A_{Y_2(\mathfrak{a})}$, $P^A_{Y_2(\mathfrak{b})}$, $P^A_{Adj}$ and $P^A_\emptyset$:
\be
P^A_{Adj}(x\to x^2)&=&-{\{t\}\{q^2\}\over\{q\}\{qt\}}\underbrace{\left(P^A_{[31^{N-3}]}+P^A_{[332^{N-3}]}\right)}_{X_2}+
\underbrace{P^A_{[42^{N-2}]}}_{Y_2(\mathfrak{a})}+{\{t\}^2\{q^2\}^2\over\{q\}^2\{qt\}^2}\underbrace{P^A_{[221^{N-4}]}}_{Y_2(\mathfrak{b})}-\nn\\
&-&{\{q^2\}^2\{t\}\{{q\over t}\}\{{q^4\over t^4}A^2\}\over
\{q\}\{{q^3\over t}A\}\{{q^2\over t^2}A\}\{{q\over t^3}A\}}\underbrace{P^A_{[21^{N-2}]}}_{Adj}
+{\left(\{{q^3\over t}A\}+2\{{q\over t}A\}-\{{q^3\over t^3}A\}\right)\{q\}\{A\}\{A/t\}\over\{t^2\}\{{q\over t}A\}\{{q^2\over t}A\}\{{q\over t^2}A\}}
\ee
A non-trivial new feature is that the adjoint representation contributes at $t\ne q$ (cf. with \cite[Eq.(28)]{MMMuniv}).
Now the hyperpolynomial of the torus knot $T[2,2n+1]$ is of the form
\be\label{knot}
\mathfrak{H}_{Adj,Adj}^{T[2,2n+1]} &\sim&
-2\mathfrak{f}_{X_2}^{-2n-1}\gamma_{X_2}{\{t\}\{q^2\}\over\{q\}\{qt\}}\MD^A(X_2)+\\
&+&\mathfrak{f}_{Y_2(\mathfrak{a})}^{-2n-1}
\gamma_{Y_2(\mathfrak{a})}\MD^A(Y_2(\mathfrak{a}))+
\mathfrak{f}_{Y_2(\mathfrak{b})}^{-2n-1}\gamma_{Y_2(\mathfrak{b})}{\{t\}^2\{q^2\}^2\over\{q\}^2\{qt\}^2}\MD^A(Y_2(\mathfrak{b}))-\nn\\
&-&\mathfrak{f}_{Adj}^{-2n-1}\gamma_{Adj}{\{q^2\}^2\{t\}\{{q\over t}\}\{{q^4\over t^4}A^2\}\over
\{q\}\{{q^3\over t}A\}\{{q^2\over t^2}A\}\{{q\over t^3}A\}}\MD^A(Adj)
+\gamma_\emptyset
{\left(\{{q^3\over t}A\}+2\{{q\over t}A\}-\{{q^3\over t^3}A\}\right))\{q\}\{A\}\{A/t\}\over\{t^2\}\{{q\over t}A\}\{{q^2\over t}A\}\{{q\over t^2}A\}}\nn
\ee
where  $\gamma$-factors have to be separately evaluated. In \cite{DMMSS}, they were found using that $\mathfrak{H}_{Adj,Adj}^{T[2,1]}$ is unknot and, hence is proportional to $\MD^A(Adj)$. It is enough strong condition, since it works at any $N$ for one and the same Young diagram, while the $\gamma$-factors remain independent of $N$. Hence, one obtains a series of conditions which sometimes allows one to fix all the $\gamma$-factors.

In the adjoint representation case, the situation is more complicated: we deal with just one Young diagram at any concrete $N$, and hence there is only one condition for six $\gamma$-factors. This makes the whole calculation of the $\gamma$-factors far more involved (see, e.g., \cite{ChE}).

\fbox{Thus the problem of universalization of $\gamma$-factors for {\it knot} polynomials remains open.}

\subsection{From non-universal framework to universal formulas}

Besides the adjoint Macdonald dimension (when one starts from non-universal Macdonald polynomials and generates the universal answer at the special point $q^{2\rho_k}$), there are two more quite interesting examples, when, starting from a non-universal framework, one can generate the universal object, the Hopf adjoint hyperpolynomial.

First of all, in the case of $A$ series, the Hopf link invariant in the adjoint representation in Chern-Simons theory can be also obtained as the 4-point correlation function of topological string on the resolved conifold with fundamental representation on the external legs \cite{AKMM1}. In order to obtain it, say, in the $D$ series case, one has to consider the orientifold \cite{Orient}. It would be interesting to understand how these different pictures give rise to universal formulas.

Note that the situation is far more involved in the refined Chern-Simons theory: in this case, the Hopf hyperpolynomial does not coincide with the 4-point function \cite{AKMM2}, since this later gives rise \cite{MMhopf} to the Hopf superpolynomial \cite{DGR} associated with the Khovanov-Rozansky approach \cite{Kh1,Kh2,Kh3,KhR,DM3}. It is still unknown if the superpolynomials are described by universal formulas.

The second example, given by the so called CMM (Cherednik-Macdonald-Mehta) formulas \cite{Che,EK,ChaE,MMP}. The l.h.s. of these formulas is an average of the product of two Macdonald polynomials for arbitrary root system, which are in no way universal quantities. The average produces the Hopf hyperpolynomial \cite{MMP}, which, in the particular case of adjoint representations, admits the universal form (\ref{Hopfu}). The average can be realized equivalently by a Gaussian integral, or by a contour integral with the $\theta$-function density. The formulas in the latter case look as follows\footnote{According to the Macdonald-Mehta theorem, the normalization of the integral is
\begin{equation}
\oint \prod_i{dx_i\over x_i}\Delta^R(x)\theta^{(+)}(x) = |W_R|\ \Delta_{+}^R (q^{2\rho_k})\nn
\end{equation}
where $|W_R|$ is order of the Weyl group $W_R$ of the root system $R$.}:
\be\label{CMM}
\left<P^R_{Adj}(x)P^R_{Adj}(x^{-1})\right>_+:=\displaystyle{\oint \prod_i{dx_i\over x_i}\Delta^R(x)P^R_{Adj}(x)P^R_{Adj}(x^{-1})\theta^{(+)}(x)
\over \oint \prod_i{dx_i\over x_i}\Delta^R(x)\theta^{(+)}(x)}=
     \mathfrak{f}_{Adj}^4\cdot
    \overline{ \mathfrak{H}}_{Adj,Adj}^{\rm Hopf}\\
\left<P^R_{Adj}(x)P^R_{Adj}(x)\right>_-:=\displaystyle{\oint \prod_i{dx_i\over x_i}\Delta^R(x)P^R_{Adj}(x)P^R_{Adj}(x)\theta^{(-)}(x)
\over \oint \prod_i{dx_i\over x_i}\Delta^R(x)\theta^{(-)}(x)}=
     \mathfrak{f}_{Adj}^4\cdot
    \overline{ \mathfrak{H}}_{Adj,Adj}^{\rm Hopf}\nn
\ee
and $t=q^k$ with an integer $k$. Here
\be
\theta^{(\pm)}(x):\ \stackrel{x_i = q^{2z_i}}{=}\ \sum_{\Lambda \in P} q^{\pm (\Lambda,\Lambda)} q^{2(\Lambda,z)}
\ee
and the sum goes over the whole weight lattice $P = \mathbb{Z}\,\omega$. Here also
\begin{equation}
    \Delta^R(x) := \Delta_{+}^R(x)\Delta^R_{-}(x),
\end{equation}
\begin{equation}
    \Delta^R_{\pm}(x):\ \stackrel{x_i = q^{2z_i}}{=}\  \prod_{\alpha \in R_{\pm}} \prod_{j=0}^{k-1} \{q^{(\alpha,z)+j}\}
\end{equation}

In fact, the r.h.s. of these formulas are proportional to the hyperpolynomial of the mirror reflected Hopf link $\overline{ \mathfrak{H}}_{Adj,Adj}^{\rm Hopf}$, which is obtained by the replace $(q,t)\to (q^{-1},t^{-1})$ in (\ref{Hopfu}).

\bigskip

Thus (\ref{CMM}) provides an interesting example of explicit formula, where the entries at the l.h.s. are absolutely non-universal, but the r.h.s., and thus the whole integral at the l.h.s. are perfectly universal. This is an interesting example of a universality ``projections" of originally non-universal quantities, which deserve better understanding.

\section{Conclusion
\label{conc}}

Following the recent analysis of \cite{BM},
we formulated the Vogel's universality conjecture for Macdonald dimensions.
To become universal, they should at least be multiplied by $(q,t)$-dependent Macdonald Littlewood-Richardson coefficients from
(\ref{main}).
We actually proved this conjecture for the simplest case of $\mu=\nu=Adj$, for the ADE series,
and posed a number of challenging questions about (validity of) its further generalizations.
Hopefully they will attract attention and the problem of consistency between Vogel universality and Macdonald theory
will be clarified, if not resolved.

To extend these conjectures further is not that simple. First of all, taking them beyond the ADE algebras looks very problematic.
Second, even for the $A$ and $D$ series, the structure of uirreps becomes complicated.
We remind that, in the universal form, they are often {\it not} described by nicely factorized formulas \cite{ManeIsaevKrivMkrt, IsaevProv, IsaevKriv,IsaevKrivProv}
with factors, linear in $\alpha,\beta,\gamma$,
for the first time, this happens for $X_3$ in the antisymmetric cube of the adjoint representation.
Factorization occurs only for particular algebras, moreover, if there are several irreps in one uirrep,
one gets a sum of factorized quantities.

It is  instructive to list the violations and subtleties of naive universality that we already encountered:

\begin{itemize}

\item{} Macdonald dimensions even from the adjoint sector {\it per se} are generally not universal,
they {\it can} become such only in combination with Macdonald deformation of the Littlewood-Richardson coefficients $C$.
This is very much in spirit of the idea that universality works for knots/links and not just for an {\it arbitrary} question
in representation theory.
\item{} Even the products $C\cdot {\rm Md}$ do not look universal beyond ADE, neither do the adjoint Macdonald dimensions \cite{BM}.
\item{} The  formula for one of the items, $\mathfrak{P}_{Adj}$, in (\ref{du1}) is rather complicated,
and, at this moment, its universality follows only from the fact that all other terms at both sides of (\ref{du1}) are universal.
This trick may not be sufficient for higher representations in adjoint sector.
\item{} Formulas for 2-strand links $[2,2n]$ are ADE universal.
However, it is still unclear if it remains true even for 2-strand knots $[2,2n+1]$,
and what guarantees the universality of the $\gamma$-factors.

\end{itemize}

To summarize, Macdonald deformation of the universality  is a challenging new problem,
which does not look neither fully stupid, nor enormously difficult.
The work on it would shed new light on the Vogel's universality, representation and knot theories
beyond quantum algebras, and will be a natural step in the current efforts to developing these theories
towards Yangians and quantum toroidal algebras.

\section*{Acknowledgements}

We are grateful to the organizers of the Workshop ``Universal description of Lie algebras, Vogel theory, applications" in Dubna (April, 2025). This work is supported by the RSF grant 23-41-00049.

\end{document}